\documentclass[12pt]{article}
\usepackage[square, numbers, sort&compress]{natbib}
\usepackage{latexsym,graphicx,amssymb,amsmath,mathrsfs}
\usepackage{setspace,bm}
\usepackage[breaklinks, colorlinks=true, pdfstartview=FitV, linkcolor=red, citecolor=blue, urlcolor=blue]{hyperref}
\usepackage[usenames]{color}
\usepackage{latexsym}
\usepackage{epstopdf}
\usepackage{mathtools}
\usepackage{url}
\usepackage{comment}
\usepackage{braket}
\usepackage{CJKutf8}
\PassOptionsToPackage{hyphens}{url}

\usepackage[usenames]{color}
\usepackage{latexsym}
\usepackage{epstopdf} 
\usepackage[normalem]{ulem}

\input epsf.tex

\newcommand{\beq}{\begin{equation}}
\newcommand{\eeq}{\end{equation}}
\newcommand{\be}{\begin{equation}}
\newcommand{\ee}{\end{equation}}
\newcommand{\bea}{\begin{eqnarray}}
\newcommand{\eea}{\end{eqnarray}}

\makeatletter

\@addtoreset{equation}{section}
\makeatother

\def\href#1#2{#2}

\begin{document}

\baselineskip=15.5pt
\pagestyle{plain}
\setcounter{page}{1}


\begin{titlepage}
\begin{flushleft}
       \hfill                       FIT HE - 24-03 \\
       \hfill                       KYUSHU-HET 135 \\
\end{flushleft}

\begin{center}
  {\huge Chiral condensate \\  \vspace*{2mm}
 in a holographic 
dilute 
nuclear matter   
}
\end{center}

\begin{center}

\vspace*{5mm}
{\large ${}^{\dagger}$Kazuo Ghoroku\footnote[1]{\tt gouroku@fit.ac.jp},
${}^{\dagger}$Kouji Kashiwa\footnote[2]{\tt kashiwa@fit.ac.jp},
${}^{\S}$Motoi Tachibana\footnote[3]{\tt motoi@cc.saga-u.ac.jp} \\
and ${}^{\ddagger}$Fumihiko Toyoda\footnote[4]{\tt ftoyoda@fuk.kindai.ac.jp}
}\\

\vspace*{2mm}
{${}^{\dagger}$Fukuoka Institute of Technology, Wajiro, 
Fukuoka 811-0295, Japan\\}
\vspace*{2mm}
{${}^{\S}$Department of Physics, Saga University, Saga 840-8502, Japan\\}
\vspace*{2mm}
{{${}^{\S}$Center for Theoretical Physics, Khazar University, 41 Mehseti Street, \\
Baku, AZ1096, Azerbaijan,\\}}
\vspace*{2mm}
{${}^{\ddagger}$Faculty of Humanity-Oriented Science and
Engineering, Kinki University,\\ Iizuka 820-8555, Japan}
\vspace*{3mm}
\end{center}

%

\begin{center}
{\large Abstract}
\end{center}

We study chiral condensate in cold nuclear matter on the basis of holographic theory, which would be dual to a baryon system in quantum chromodynamics (QCD) in the confinement phase.
Our model is a holographic model based on the D$3$/D$7$ branes.
The magnitude of the chiral condensate obtained in our model is found to gradually increase with increasing baryon density $n$ within the range of the dilute nucleon gas, $n<1$, where our model is available.
This is a result for the chiral condensate in the holographic model as a function of $n$, which can be compared to the chiral perturbation theory.

\noindent

\vfill
\begin{flushleft}
\end{flushleft}

\end{titlepage}

\newpage

\vspace{1cm}


\section{Introduction}

It is important to study the behavior of the chiral condensates in quantum chromodynamics (QCD) to understand the dynamical properties dominated by Yang-Mills theory in the nonperturbative regime.
In particular, several interesting properties of QCD are expected to appear in the nonperturbative regime.
The AdS/CFT duality is very useful in approaching this kind of dynamics \cite{Maldacena:1997re,Gubser:1998bc,Witten:1998qj,Erdmenger:2007vj,Ghoroku:2013gja,Ghoroku:2021fos}.

We propose a holographic model that is supposed to be dual to low-temperature baryonic matter \cite{Ghoroku:2013gja,Ghoroku:2021fos,Sakai:2004cn,Hata:2007mb,Hashimoto:2008zw}.
Using this model, we can calculate the physical quantities in baryonic matter. 
The calculation is performed in the confinement phase at low temperature and small baryon number density ($n$).
Our model is based on the type IIB theory, and the $N_f (=2)$ flavor probe D$7$ branes are introduced. 
Then, the profile of the brane and the instantons for the non-Abelian flavor gauge-vector configurations are appropriately set.
\footnote{We notice that the procedure is parallel to Ref.\,\cite{Ghoroku:2013gja}, where IIA model is discussed.}
In the model, the Chern-Simons term, which is obtained as a coupling term between the vector mesons, is important in order to couple instantons to the baryon charge density.

According to the model mentioned above, various physical quantities are obtained by solving the equation of motion (EOM) of bulk fields set appropriately through our holographic effective action for probes. 
Then we can obtain the chiral condensate $C=\langle \bar{q}q\rangle$ at some definite baryon number density $n$, which is set by identifying the instanton as a baryon.
The results provide us with the relation between the chiral condensate and $n$. 
However, in order to understand the correct $n$-dependence of the chiral condensate, we must also be careful of the size of the baryon. 
The size of the baryon is also an important parameter that controls the chiral condensate.
However, the size is not an independent parameter, but it depends on $n$. 
Then, the $n$-dependence of the magnitude of the chiral condensate is not monotonic, and it may have an interesting structure. 
This point will be ensured by the analysis in the high density region, where the dilute gas approximation is, however, not useful. 
This point will be discussed in future work.

In the next section and Sec.\,\ref{sec:3}, a holographic model is given for a baryon system. 
In Sec.\,\ref{sec:4} and \ref{sec:5}, EOM of the D7 profile and a $U(1)$ bulk gauge field, which is dual to the baryon charge current, are solved in the low-temperature confinement phase. 
The summary and discussion are given in the last section.

\section{D3/D7 model for confining YM theory}

The expectation value of QCD operators in the confinement phase can be examined in terms of a holographic model dual to a nuclear matter. 
The model proposed here is based on the $10$d IIB theory retaining the dilaton $\Phi$, axion $\chi$ and self-dual five-form field strength $F_{(5)}$.
Under the Freund-Rubin ansatz for $F_{(5)}$, $F_{\mu_1\cdots\mu_5}=-\sqrt{\Lambda}/2~\epsilon_{\mu_1\cdots\mu_5}$ \cite{Kehagias:1999iy,Liu:1999fc}, and for the $10$d metric as $M_5\times S^5$ or $ds^2=g_{MN}dx^Mdx^N+g_{ij}dx^idx^j$, we have found interesting solutions.

The five dimensional $M_5$ part of the
solution is obtained by solving the following reduced $5$d action,
\begin{align}
    S = \frac{1}{2\kappa^2} \int d^5x \sqrt{-g} 
        \Bigl( R + 3\Lambda - \frac{1}{2}(\partial \Phi)^2
                 + \frac{1}{2}e^{2\Phi} (\partial \chi)^2
        \Bigr),
\label{5d-action}
\end{align}
which is written in the Einstein frame and we set as $\alpha'=g_s=1$. 
%
The solution for the metric is expressed as
\begin{align}
    ds^2_{10}
    &= G_{MN}dX^{M}dX^{N}
    \nonumber\\
    &=e^{\Phi/2}
    \Bigl[
            {r^2 \over R^2}A^2(r) \Bigl\{ -dt^2+(dx^i)^2 \Bigr\}+
            \frac{R^2}{r^2} dr^2+R^2 d\Omega_5^2
    \Bigr],
\label{finite-c-sol}
\end{align}
where $M,~N=0,..., 9$ and $R=\sqrt{\Lambda}/2=(4 \pi N)^{1/4}$.

The supersymmetric solution is obtained as
\begin{align}
    A = 1,~~~~ e^\Phi &= 1 + \frac{q}{r^4}, 
\label{dilaton}    
\end{align}
with
\begin{align}
     \chi = -e^{-\Phi} + \chi_0,
\end{align}
where $q$ represents the vacuum expectation value (VEV) of the gauge-field condensate~\cite{Ghoroku:2004sp}. 
In this configuration, the four-dimensional boundary represents the 
$\cal{N}$=2 supersymmetric Yang-Mills (SYM) theory. 
In this model, we find quark confinement in the sense that we find a linear rising potential between quark and anti-quark with the tension $\sqrt{q}/R^2$ \cite{Kehagias:1999iy,Ghoroku:2004sp}.

For the non-supersymmetric case, the solution is given by Eq.\,(\ref{finite-c-sol})
and
\begin{align}
    A(r) = \Bigl[ 1 - \Bigl( \frac{r_0}{r} \Bigr)^8 \Bigr]^{1/4},~~~~
    e^{\Phi} = \Bigl[ \frac{(r/r_0)^4+1}{(r/r_0)^4-1} \Bigr]^{\sqrt{3/2}},\qquad
\chi=0\, .
\label{non-susy-sol}
\end{align}
This configuration has a singularity at the horizon $r=r_0$. 
So, we cannot extend our analysis to the region near this horizon, where higher curvature contributions are important.
This theory provides confinement and chiral symmetry breaking.
The latter means that we find a non-zero chiral condensate for the massless quark.
In other words, in this theory, a dynamical quark mass would be generated for a massless quark.
This point is different from the above supersymmetric background solution.
The confinement is sustained by the gauge condensate, which is proportional to $r_0^4$ in the present case \footnote{This point is easily assured by expanding $e^{\Phi}$ in Eq.\,(\ref{non-susy-sol}) by the powers of $r_0/r$.}, as in the supersymmetric case.

\section{\bf D$7$ brane embedding and chiral condensate}~
\label{sec:3}
The D$7$ quark-brane is embedded as a probe on the background defined by Eqs.\,(\ref{finite-c-sol}) and (\ref{non-susy-sol}). 
Here, the extra six-dimensional part in the metric (\ref{finite-c-sol}) can be written as
\beq
 \frac{R^2}{r^2} dr^2 + R^2 d\Omega_5^2
 = \frac{R^2}{r^2}
   \Bigl[ d\rho^2+\rho^2d\Omega_3^2+(dX^8)^2+(dX^9)^2 \Bigr],
\eeq
where $r^2=\rho^2+(X^8)^2+(X^9)^2$.
Then, we obtain the induced metric for the D7 brane as
\begin{align}
    ds^2_8
    = e^{\Phi/2}
      \Bigl[
            {r^2 \over R^2} A^2 \Bigl\{ -dt^2+(dx^i)^2 \Bigr\}
           + \frac{R^2}{r^2} \Bigl\{ \Bigl( 1 + (\partial_{\rho}w)^2 \Bigr) d\rho^2
                                   + \rho^2 d\Omega_3^2 \Bigr\}
      \Bigr],
\label{D7-metric}
\end{align}
where we set as $X^8=w(\rho)$ and $X^9=0$ without loss of generality due to the rotational invariance in the $X^8$-$X^9$ plane. 
Namely, $X^8$ is chosen as the field being dual to the chiral condensate.

\subsection{DBI action with instantons}

Here, we consider the case of stacked two D$7$-probe branes, and the action is given by a Dirac-Born-Infeld (DBI) action according to Refs.\,\cite{Erdmenger:2007vj,Ghoroku:2013gja,Ghoroku:2021fos} as
\begin{align}
      S_{D_7} = -\tau_7\int d^{8}\xi \, e^{-\Phi} \, {\rm Str}\, L ,
\end{align}
where ${\rm Str}$ denotes the symmetric trace of flavor $U(2)$ and
\begin{align}
      L &= \sqrt{-{\rm det}\left(f_0+f_1\right)},~~~~
     (f_0)_{ab} = \Bigl( {\cal G}_{ab} + B_{ab} + \tilde{F}_{ab} \Bigr) \tau_0, 
\end{align}
here
\begin{align}
  {\cal G}_{ab} &= G_{MN}\partial_a X^M\partial_b X^N \,, \\
  B_{ab} &= B_{MN}\partial_a X^M\partial_b X^N=-B_{ba}\,, \\
  \tilde{F}_{ab} &= 2\pi\alpha'F_{ab}\, , \\
  {(f_1)}_{ab} &= 2\pi\alpha' F_{ab}^i\tau_i\, ,
\end{align}
with $(a,b=0, 1, \cdots, 7)$, $(M,N=0, 1, \cdots, 9)$, the unit matrix $\tau_0$ and Pauli's spin matrices $\tau_i$.
The Str part is expanded as
\begin{align}
  {\rm Str} \, L &= {\rm Str} \Bigl\{ \sqrt{-{\rm det}(f_0)}
                          \Bigl( 1 - \frac{1}{4} {\rm tr} \, x^2
                                   + \frac{1}{8} ({\rm tr} \, x)^2 + \cdots \Bigr)
                          \Bigr\} ,
\end{align}
with
\begin{align}
    x &= f_0^{-1} f_1,
\end{align}
where ${\rm tr}$ denotes the trace of the coordinate index. The linear term vanishes for ${\rm Str}$, then it is dropped.
The coordinates are set as 
\begin{align}
    (\xi^0,\xi^1,\xi^2,\xi^3,\xi^4,\cdots) = (x^0,x^1,x^2,x^3, \rho,\cdots), 
\end{align}
and we make the ansatz for the $SU(2)$ bulk gauge fields as the instanton, then we write  
\begin{align}
   \tilde{F}_{ab} &= 2\partial_{[a}A_{b]}, \\
   B_{ab} &=0, \\
   A_b &= A_b(\rho)\delta_b^0, \\
   (f_1)_{ij} &= Q(x^m-a^m,\sigma)\epsilon_{ija}\tau^a,
   \label{an1} \\
   (f_1)_{iz} &= Q(x^m-a^m,\sigma)\tau^i,
   \label{an2}
\end{align}
where 
\begin{align}
    Q &={ \frac{2\sigma^2}{[(x^m-a^m)^2+\sigma^2]^2},}
\end{align}
here $\sigma$ is the instanton size, $\epsilon_{123z}=1$, $i,j=1,2,3$ and {$m=(1,\dots ,4)$ with $x^4=\rho$.}

\vspace{.3cm}
Then the above series of $
f_1$, by retaining its lowest order, are obtained as
\begin{align}
    {\rm Str} \, L
    &= 2\sqrt{-{\rm det}(f_0)} \Bigl[  1+{3\over 2}Q^2
       \Bigl\{ \Bigl( {\cal G}^{11} \Bigr)^2
       + {\cal G}^{11} {\cal G}^{zz} \Bigr\} \Bigr], \\
      {\cal G}_{ab} &= G_{MN}\partial_a X^M\partial_b X^N.
      \label{Inst-Contri}
\end{align}
Hereafter, we neglect the higher order terms of the non-Abelian gauge fields.
{In the equations,} $F_{ab}=\partial_aA_b-\partial_bA_a$.
${\cal G}_{ab}= \partial_{\xi^a} X^M\partial_{\xi^b} X^N G_{MN}~(a,~b=0,\ldots, 7)$
and $\tau_7=[(2\pi)^7g_s~\alpha'~^4]^{-1}$ represent the induced metric and
the tension of D$7$ brane, respectively.

By taking the canonical gauge, we arrive at the following D$7$ brane
action;
\begin{align}
    S_{\rm D7} &= -4 \pi^2 \tau_7 \int d^4x d{\rho}\,\, L_7 Q_1,
    \label{D7-1}
\end{align}
where
\begin{align}
    L_7 &= B \sqrt{A^2 [1+(w')^2] - (\tilde{A}_0')^2  e^{-\Phi}},
    \label{eq-L7} \\
    Q_1 &= 1+{3\over 2} n \bar{q}_0^2 g_0, \\
   \bar{q}_0^2 &= \frac{9}{8} \frac{\pi^2 \sigma^4}{(\rho^2+\sigma^2)^{5/2}}, \\
   g_0 &= e^{-\Phi} \Bigl[ \frac{R^4}{r^4 A^4} + \frac{1}{A^2 \{ 1+(w')^2 \}} \Bigr],
\end{align}
here $\tilde{A}_{0}=2\pi\alpha'{A}_{0}$,
\begin{align}
 n = \frac{N_I}{V_3},
 \label{eq:n_I}
\end{align}
denotes the instanton density, and
\begin{align}
   B = \rho^3 A^3 e^{\Phi}.
   \label{eq-B}
\end{align}


\subsection{CS term} 
To construct a reduced $5$D action, we consider the following form of CS term for $N_f=2$ \cite{Hata:2007mb,Ghoroku:2013gja},
\begin{align}
    S_{CS}
    &= \kappa_{CS} \epsilon^{m_1\cdots m_4} \int d^3\theta
        dx^{\mu}d\rho \, \frac{3}{2} f_3(\theta) A_0(\rho)
                         {\rm Tr} (F_{m_1m_2}F_{m_3m_4}),
\end{align}
where $f_3(\theta)d^3 \theta = dC_2$.
For the instanton configuration, we have
\begin{align}
    S_{CS}
    &= 36 \kappa_{CS} N_0 \int dx^{\mu} d\rho \, A_0
       \frac{n\, \pi^2\sigma^4}{(\rho^2+\sigma^2)^{5/2}},
    \label{CS} 
\end{align}
where $N_0=\int d^3\theta f_3(\theta) $.
This term is then induced in our calculation as a coupling of $A_0$ and instantons.


\section{Effective probe action and chiral condensate} 
\label{sec:4}

The effective probe action to be solved is given by adding Eqs.\,(\ref{D7-1}) and (\ref{CS}) as
\begin{align}
    S_{\rm eff}
    &= S_{D7} + S_{CS}
      = -4\pi^2 \tau_7 \int d^4x d\rho \left (L_7 Q_1
             -\tilde{\kappa}_{CS}A_0 n {\pi^2\sigma^4 \over (\rho^2+\sigma^2)^{5/2}}\right),
    \label{multi-2} \\
    \tilde{\kappa}_{CS} &= \frac{36 N_0}{4\pi^2\tau_7} \kappa_{CS},  
    \label{multi-1}
\end{align}
where $n$ is the density of instantons defined in Eq.\,(\ref{eq:n_I}).

\vspace{.3cm}
\noindent{\bf Equations of motion}

\begin{figure}[t]
\vspace{.3cm}
\begin{center}
\includegraphics[width=7cm]{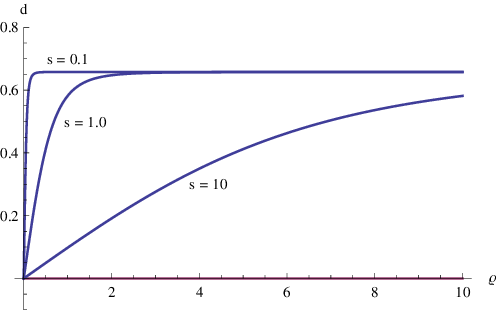}
\includegraphics[width=7cm]{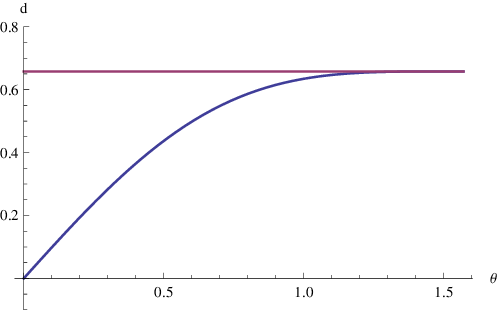}
\caption{\small 
The left panel shows the $\rho$-dependence of $d$ for some value of $\sigma=s$. 
The right panel is the extended figure of the left panel near the origin.
The horizontal line on the right panel shows the value of $d(\infty)=2 n \tilde{\kappa} \pi^2/3=2\bar{d}$ where $\bar{d}$ is given by Eq.\,(\ref{asymp-d}).
}
\label{rho_d}
\end{center}
\end{figure}

From the above effective action (\ref{multi-2}), we obtain the EOMs of $w(\rho)$ and $A_0(\rho)$ as
\begin{align}
    & -\partial_{\rho}
      \left[ B \frac{A^2w'}{\sqrt{A^2 \{ 1+(w')^2 \} - (\tilde{A}_0')^2 e^{-\Phi}}} Q_1 \right]
           + \frac{w}{r} \partial_r ( L_7 \, Q_1 ) 
    \nonumber\\
    &= -\frac{3}{2} n \partial_{\rho}
       \left[ \bar{q}_0^2 e^{-\Phi} {2 w' \over A^2 \{1+(w')^2\}^2 } L_7 \right],
       \label{eq-w-I}
\end{align}
\begin{align}
   \partial_{\rho}
   \left( B \frac{\tilde{A}_0' e^{-\Phi}}
                 {\sqrt{A^2 \{ 1+(w')^2 \} - (\tilde{A}_0')^2  e^{-\Phi}}} Q_1 \right)
    = \frac{8}{9} \bar{q}_0^2 \tilde{\kappa} n.
    \label{A0I}
\end{align}
Equation (\ref{A0I}) is rewritten by integrating over $\rho$, and we have
\begin{align}
    B \frac{\tilde{A}_0' e^{-\Phi}}{\sqrt{A^2(1+(w')^2)-(\tilde{A}_0')^2  e^{-\Phi}}} Q_1 = d(\rho),
    \label{d-rho}
\end{align}
where
\begin{align}
    d(\rho)
    &= \frac{8}{9} \tilde{\kappa} n \int_0^{\rho} d\rho \, \bar{q}_0^2,
    \label{dr-1} \\
    &= n \tilde{\kappa} \pi^2
       \left( \frac{1}{12} \sin(3\theta) + \frac{3}{4} \sin (\theta) \right),
       \label{dr-2} \\
    \tan(\theta) &= \frac{\rho}{\sigma}.
     \label{dr-3} 
\end{align}
The $\rho$-dependence of $d$ with fixed $\sigma$ is shown in Fig.\,\ref{rho_d}.

\vspace{1cm}
\noindent {\bf Chemical potential and charge density}

The meaning of $d(\rho)$ is found by the asymptotic expansion of the Eq.\,(\ref{d-rho}) 
according to the holographic ansatz,
\begin{align}
   A_0 &= \mu - \frac{\bar{d}}{\rho^2} + \cdots\,\,\,  (\rho\rightarrow \infty), 
\end{align}
where $\mu$ and $\bar{d}$ represent the chemical potential and the baryon number density of the dual theory with a nuclear matter, respectively. 
Picking up the leading term with respect to the power of $\rho$, we found
\begin{align}
    \bar{d} = \frac{n \tilde{\kappa} \pi^2}{3}.
    \label{asymp-d}
\end{align}

For the chemical potential $\mu$, it is obtained according to the following procedure. First, $A_0'$ is written using Eq.\,(\ref{d-rho}) as 
\begin{align}
    A_0' = A e^{\Phi/2} d\sqrt{ \frac{1+(w')^2}{d^2+B^2Q_1^2e^{-\Phi}} },
   \label{A0-eq}
\end{align}
Then, substituting this form of $A_0'$ into Eq.\,(\ref{eq-w-I}), we can solve it to obtain $w(\rho)$. 
From this solution $w(\rho)$, we find the chiral condensate $C$ and $m_q$ and the explicit form of $A_0'$, which provides the chemical potential as
\begin{align}
  \mu=\int_0^{\infty}d\rho Ae^{\Phi/2} d\sqrt{{ 1+(w')^2 \over d^2+B^2Q_1^2e^{-\Phi}}} \,\,.
\end{align}
The solution $w(\rho)$ is given as a solution of the other EOM (\ref{eq-w-I}) as shown in the next section.

\section {Chiral condensate and baryon density}
\label{sec:5}

Now it is possible to solve Eq.\,(\ref{eq-w-I}) with $A_0'$ and $d(\rho)$ given above, then we obtain the solution $w(\rho)$, which provides us with information about the relation between the baryon number density $n$ and the chiral condensate. 
 
The solutions of Eq.\,(\ref{eq-w-I}) are obtained by imposing the condition
\begin{align}
   w'(0) = 0,
   \label{bound1}
\end{align}
and $w(0)~(>r_0)$ at $\rho =0$. 
These conditions are consistent with the fact that the profile function should be smooth at $\rho=0$ in the four-dimensional manifold in which the instantons are embedded. 

{Then, using the presumed asymptotic behavior of $w$;}
\begin{align}
    w = m_q + \frac{C}{\rho^2} + \cdots\,\,\,  (\rho\rightarrow \infty),
    \label{asym-w}
\end{align}
we can read the values of $m_q$ and the chiral condensate $C$ from the solutions. 
This procedure is repeated by changing the value of $n$. In this case,
$w(0)$ is changed at the same time to keep the value of $m_q$ for the previous $n$. 
This procedure is 
{repeated} by changing $n$, and we obtain $C$ as a function of $n$ for a fixed $m_q$ and $\sigma$. 




\vspace{0.5cm}
\subsection { Small $n$ expansion}\label{5.1}

\begin{figure}[htbp]
\vspace{.3cm}
\begin{center}
\includegraphics[width=8cm]{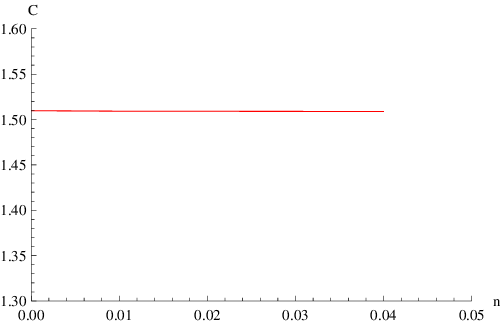}
\caption{\small The chiral condensate $C$ versus baryon density $n$ for $m_q\sim 0$.
Here, $R=1$, $r_0=1$, $\kappa=1,$ and $\sigma=1$.
}
\label{mqC}
\end{center}
\end{figure}

Plugging $A_0'$, which is given by Eq.\,(\ref{A0-eq}), in Eq.\,(\ref{eq-w-I}), we can obtain the solution $w(\rho)$, whose asymptotic form provides us with the chiral condensate $C$ and the quark mass $m_q$ as in Eq.\,(\ref{asym-w}).
One of our purposes is to investigate how $C$ depends on the baryon number density $n$.
Since Eq.\,(\ref{eq-w-I}) is given for the dilute instanton gas approximation, we investigate the relationship between $C$ and $n$ in the small $n$ region, where the dilute gas approximation may be useful. 
In this case, it is enough to retain the linear term of $n$ in the equation as the lowest approximation.
In this procedure, $A_0'$ is treated as linear to $n$ as implied from Eq.\,(\ref{A0-eq}), and $A_0'$ appears as $(A_0')^2$ in the equation. 
Then, this term does not contribute to the linear $n$ approximation.
And we have
\begin{align}
    - \partial_{\rho} \left( \frac{A B w'}{\sqrt{1+(w')^2}} \right)
    + \frac{w}{r} \sqrt{ 1 + (w')^2 } \partial_r \left( AB \right) 
    =  n F_1 (w,w',\rho) + O(n^2), 
\end{align}
with
\begin{align}
     F_1(w,w',\rho)
     &= \partial_{\rho} \left( \frac{A B w'}{\sqrt{1+(w')^2}}
                               \frac{3}{2} \bar{q_0}^2 g_0 \right)
      - \frac{w}{r} \sqrt{ 1 + (w')^2 } \partial_r
        \left( AB \frac{3}{2} \bar{q_0}^2 g_0 \right)
      \nonumber \\
      &  -\frac{3}{2} \partial_{\rho} \left( \frac{2 B w'}{\{1+(w')^2\}^{3/2}}
                                     \bar{q_0}^2  e^{-\Phi} \right).
\end{align}

For $n=0$, we find a solution of $w$ which shows that the chiral symmetry is spontaneously broken.
Here, we solve the equation in which the linear terms of $n$ are included, then examine how the chiral condensate is controlled by the baryon density $n$.  An example of this estimation for a set of parameters of the theory used here is shown in Fig.\,\ref{mqC} for $m_q=0$. 
Within the approximation, $C$ is almost unchanged with increasing $n$.

\vspace{.5cm}
\subsection { Solution for full form of EOM }
\label{5.2}

\begin{figure}[htbp]
\vspace{.3cm}
\begin{center}
\includegraphics[width=10cm]{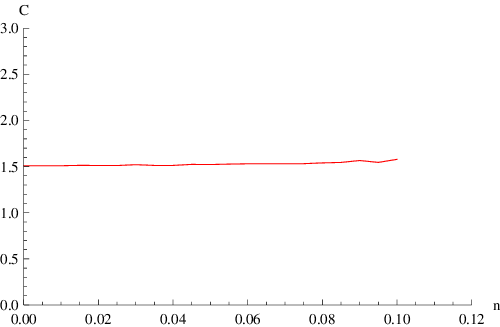}
\caption{\small The chiral condensate $C$ versus the baryon density $n$ for $m_q = 0$.
Here, $R=1, r_0=1, \kappa=1,$ and $\sigma=1$. The values of $C$ are shown for $n\leq 0.10$. 
}
\label{mqC1}
\end{center}
\end{figure}

\begin{figure}[htbp]
\vspace{.3cm}
\begin{center}
\includegraphics[width=12cm]{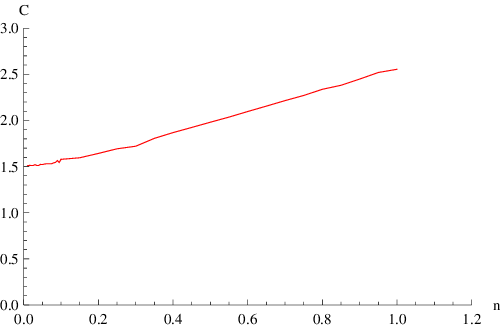}
\caption{\small The chiral condensate $C$ versus baryon density $n$ for $m_q = 0$ and $n\leq 1.0$.
Here, $R=1, r_0=1, \kappa=1,$ and $\sigma=1$.
}
\label{mqC12}
\end{center}
\end{figure}

Next, we solve Eq.\,(\ref{eq-w-I}) without any approximation. Under the same setting of the parameters given above, the $C$-$n$ relations are obtained and shown in Figs.\,\ref{mqC1} and \ref{mqC12}. 
It should be noted that here we neglect the $n$ dependence of $\sigma$ as the approximation performed in Sec.\,\ref{5.1}. 
To see the qualitative behavior, we set $\sigma=1$ as in Sec.\,\ref{5.1} to compare the resultant $C$ with the one given in the previous section.
However, we observe that $C$ slowly increases with $n$ for $n>0.1$ in Fig.\,\ref{mqC1}, and this behavior continues to $n=1.0$ as shown in Fig.\,\ref{mqC12}. 
{Because of the numerical error coming from the solving EOMs, there is a  small non-monotonic behavior of $C$ at small $n$.
This is not a physical behavior and will be removed when we improve the numerical accuracy.}

This observation indicates that we must be aware when we evaluate the approximated result obtained by neglecting the higher-order terms of $n$ even for very small $n$. 
{In addition,} we should notice that the behavior of increasing $C$ with $n$ is seen for all ranges of $n$ where the dilute baryon gas approximation is available.


We should notice that the instanton size $\sigma$ is set here as a constant.
However, $\sigma$ must be defined {using the procedure of minimization of the} free energy of the system \cite{Ghoroku:2021fos}.
Then, $\sigma$ has the $n$ dependence, and this $n$ dependent $\sigma$ will lead to a more correct $n$ dependent $C$.
However, such an investigation remains for future work here.



\newpage


\subsection{Comparison with approaches from 4D nuclear theory}

Up to now, we have shown our results for the chiral condensate at finite baryon density based on the holographic approach.
{Here, let us see} another approach based on quantum field theory and make some comparisons.

Such an approach is the chiral perturbation theory (ChPT). 
This is an effective field theory  constructed with a Lagrangian consistent with the underlying symmetry, i.e. the chiral symmetry of QCD. The dominant degrees of freedom of ChPT are pions so that the Lagrangian is described in terms of the pions, the masses and the low energy constants which are fixed by the experiments. ChPT has been applied to the finite temperature and baryon density and the chiral condensate has been evaluated. 

In Ref.\,\cite{Goda:2013bka}, for example, the authors evaluated the chiral condensate at finite baryon density within the linear density approximation
{.Then, it is shown} that the value of the chiral condensate is reduced as the baryon density increases. This is compared with our result of Sec.\,\ref{5.1}, where we performed the small $n$ expansion to the equations of motion to obtain 
{an almost constant behavior of} the chiral condensate (Fig.\,\ref{mqC}). 
If the chiral condensate decreases as the baryon density increases, the masses of hadrons would drop in nuclear matter through the QCD sum rules \cite{Hatsuda:1991ez}. 

{On the other hand,} in Ref.\,\cite{Kaiser:2007nv}, the authors have performed a treatment beyond the linear density approximation, which was done by dealing with the density-density correlations.
As a result, depending on the pion mass as well as the hadron channels, different behaviors of the chiral condensates are obtained. 
In some cases, the chiral condensate decreases up to a certain baryon density, then starts to increase at high-density region. This is {compared} to what we have found in the full calculation of Sec.\,\ref{5.2} (Fig.\,\ref{mqC1}). 
Applications of ChPT in finite temperature and baryon density have been currently developed, so we hope for future progress.


\section{Summary and discussions}



 The important point of the present model is how to add the baryon system which affects the chiral condensate. 
In the present model proposed here, we see $d(0)=0$. On the other hand, for the finite $d(0)$, the electric displacement at the end of {the D7 brane is finite and must be absorbed by some source located outside} of the D7 brane.
In order to realize this situation, a consistent coupling of the source and D7 is necessary to find an appropriate boundary condition at $\rho=0$. 
It will change the condition $w'(0)=0$ used here to solve the EOM of $w$ \cite{Evans:2012cx}.

Anyway we can say within a dilute gas approximation that the chiral condensate $C$ increases monotonically with the baryon number density in the holographic cold nuclear matter.
The meaning of this observation will be understood more deeply through several hadronic quantities related to the chiral condensate $C$ in future work.

In addition, of course it is quite difficult, we should consider the back-reaction of the matter content to the 
{geometric back-ground configuration to obtain more information of QCD with a finite baryon density}
via the holographic model.
In this sense, this study is a first step in understanding the behavior of the chiral condensate at finite density.


\section*{Acknowledgments}
We would like to appreciate Y. Nakano and D. Jido for the fruitful discussions.
{This work is supported in part by Grants-in-Aid for Scientific Research from JSPS (JP22H05112).}

\bibliographystyle{aipnum4-1}
\bibliography{ref}

\end{document}